\documentclass[%
 reprint,
 amsmath,amssymb,
 aps,
]{revtex4-1}

\usepackage{graphicx}% Include figure files
\usepackage{dcolumn}% Align table columns on decimal point
\usepackage{bm}% bold math
\usepackage[T1]{fontenc}
\usepackage{cancel}

\usepackage{graphics}
\graphicspath{{Figures/}}

\begin{document}

\preprint{APS/123-QED}

\title{Backward phase-matched second-harmonic generation from stacked metasurfaces}

\author{Timo Stolt$^{1,\#}$, Jeonghyun Kim$^{2,\#}$, S\'ebastien H\'eron$^{3}$,  Anna Vesala$^{1}$, Mikko.~J.~Huttunen$^{1}$, Robert Czaplicki$^{4}$, Martti Kauranen$^{1}$, Junsuk Rho$^{2,5,6}$ and Patrice Genevet$^{3}$}
\address{
    $^{1}$Photonics~Laboratory, Physics~Unit, Tampere~University, FI-33014~Tampere, Finland \\
        $^{2}$Department of Mechanical Engineering, Pohang University of Science and Technology (POSTECH), Pohang 37673, Republic of Korea \\
    $^{3}$Universit\'e C\^{o}te d'Azur, CNRS, CRHEA, rue Bernard Gregory, Sophia Antipolis, 06560 Valbonne, France
 \\
    $^{4}$Institute of Physics, Faculty of Physics, Astronomy, and Informatics, Nicolaus Copernicus University, Grudzi\k{a}dzka 5/7, 87-100 Toru\'n, Poland\\
    $^{5}$Department of Chemical Engineering, Pohang University of Science and Technology (POSTECH), Pohang 37673, Republic of Korea\\
    $^{6}$National Institute of Nanomaterials Technology, Pohang 37673, Republic of Korea\\
    $^{\#}$These authors contributed equally to this work.\\
}

\date{\today}

\begin{abstract}
We demonstrate phase-matched second-harmonic generation (SHG) from three-dimensional metamaterials consisting of stacked metasurfaces. To achieve phase matching, we utilize a novel mechanism based on phase engineering of the metasurfaces at the interacting wavelengths, facilitating phase-matched SHG in the unconventional backward direction. By stacking up to five metasurfaces, we obtain the expected factor of 25 enhancement in SHG efficiency. Our results motivate further investigations to achieve higher conversion efficiencies also with more complex wavefronts.

\begin{description}
\item[DOI]
\end{description}
\end{abstract}

\pacs{Valid PACS appear here}% PACS, the Physics and Astronomy
\maketitle
 
\noindent
Optical metamaterials and metasurfaces are artificial structures consisting of sub-wavelength building blocks, known as meta-atoms, and are associated with optical properties not found in nature~\cite{Soukoulis2011}. These properties include magnetism at optical frequencies, strong optical activity, negative index of refraction, and epsilon-near-zero behavior~\cite{Klein2006, Alu2007, Zhang2009}. In addition, recent work on phase-engineered metasurfaces has demonstrated the interesting possibilities to realize flat optical components, such as lenses, holographic components, and polarizers~\cite{Yu2014,Yin2017,Genevet2017, Huang2018, Arbabi2018, Ren2019,Rubin2019}.  

\begin{figure}[htbp]
\centering
{\includegraphics[]{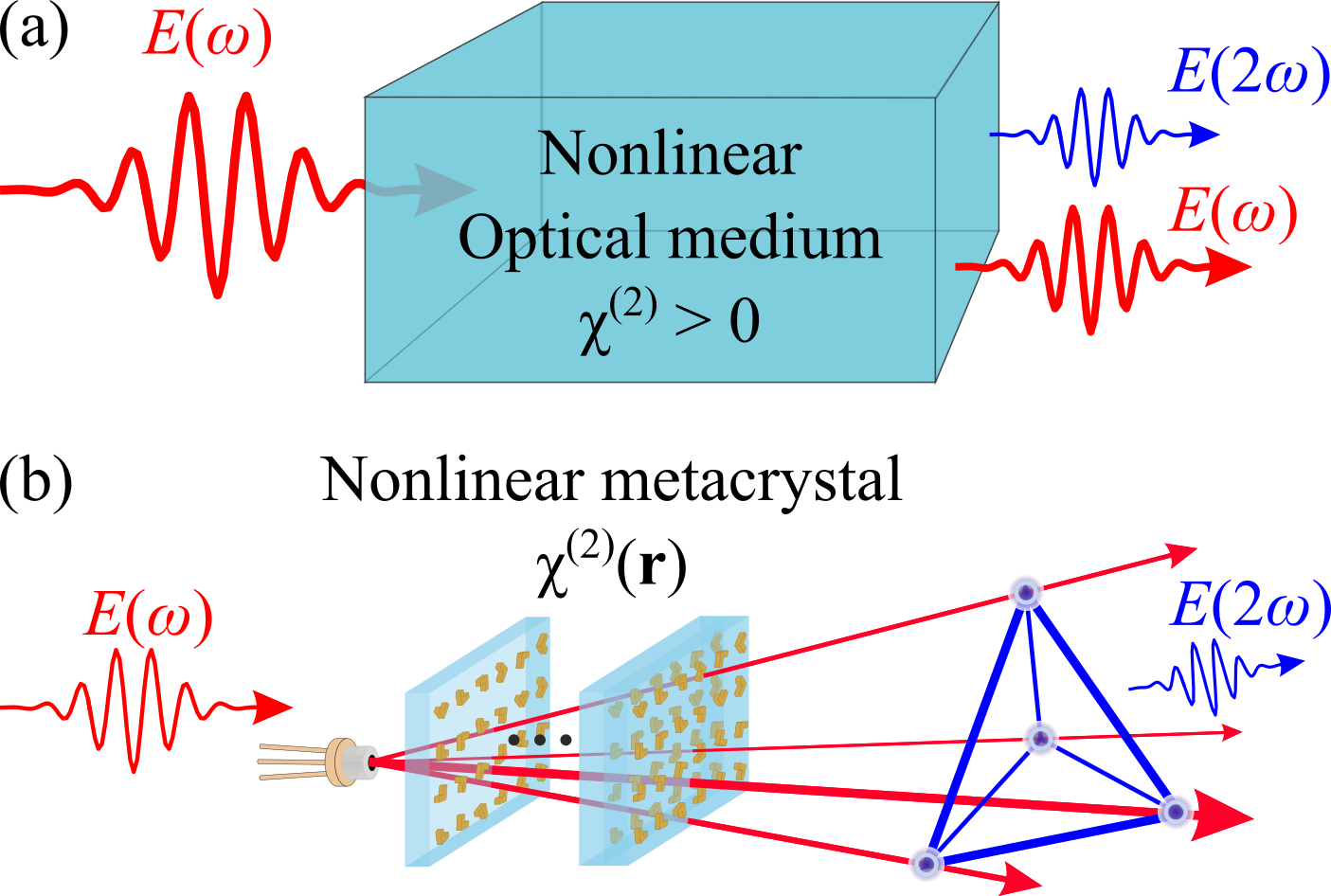}}%width=0.99\linewidth
\caption{(a) For traditional nonlinear materials, achieving phase matching and strong nonlinear responses is very restricted in terms of, e.g., the selected material and the polarization of the interacting fields. (b) Nonlinear optical processes can be phase matched with metamaterials that induce arbitrary phase changes in the interacting fields.}
\label{fig:Intro}
\end{figure}

In addition to the linear optical properties of metamaterials, their nonlinear optical responses are also becoming important. Several technologically relevant photonic applications rely on the nonlinear responses of materials, including second-harmonic generation (SHG), photon-pair generation, all-optical switching,  frequency combs, and supercontinuum generation~\cite{Kwiat1995, Brabec2000intense, Kippenberg2011comb, Shcherbakov2015}. The challenging part in these nonlinear applications is the fact that nonlinear optical processes in materials are intrinsically weak. Because of this fact, nonlinear processes in conventional materials, such as in crystals, rely on the concept of phase matching. In phase-matched materials, the generated nonlinear signal scales quadratically on the propagation length resulting in practical conversion efficiencies with sufficiently long materials (see Fig.~\ref{fig:Intro}a)~\cite{BoydBook2020, Wang2017}. 

For homogeneous materials and forward SHG signals, phase matching can be achieved if the refractive indices at the fundamental and second-harmonic frequencies are equal. However, this requirement is a significant limitation because of refractive-index dispersion, which can be overcome by the concept of quasi-phase-matching, i.e., by structuring the material in such a way that the sign of the nonlinear susceptibility is periodically reversed \cite{Lim1989}. In principle, quasi-phase-matching is a very general concept that allows any nonlinear signal to be optimized. Unfortunately, quasi-phase-matching and other traditional phase-matching schemes seem unfeasible for miniaturization of optical devices. Additionally, these techniques are restricted in terms of, e.g., polarization and the spatial profiles of the interacting waves. These limitations motivate the ongoing development of efficient and less restricted nanoscale devices.

%%%%%%%%%%%%%%%%%%%%
%%%%%%%%%%%%%%%%%%%%

Utilization of metal nanoantennas has recently emerged as a promising route towards more efficient nonlinear metamaterials~\cite{Kauranen2012review, Lee2014}. Metal nanoantennas support collective oscillations of conduction electrons, known as localized surface plasmons.
Under resonant conditions, these oscillations give rise to localized surface plasmon resonances (LSPRs), which can considerably enhance the local field near the particles~\cite{Maier2007}. Because nonlinear processes scale with high powers of the local field, the plasmon-assisted field enhancement can result in a dramatic increase in the otherwise weak nonlinear response. Consequently, numerous investigations have been carried out during the past decade in order to understand the nonlinear response of plasmonic nanoantennas~\cite{Lapine2014, ButetReview2015,  Li2017, Rahimi2018, Huttunen2019review}. So far, work on nonlinear metamaterials has focused on single planar metasurfaces limiting the achieved efficiencies. A viable route to improve the efficiencies would be to stack several metasurfaces on top of each other giving rise to phase-matching issues. In addition, such nonlinear metamaterials could provide novel capabilities to conventional phase-matching techniques relying on the intrinsic material dispersion. Particularly, use of metamaterials could allow to design phase-matched devices exhibiting arbitrary transverse phase profiles, providing therefore interesting possibilities to fabricate nonlinear metalenses and holography (see Fig.~\ref{fig:Intro}b)~\cite{Li2015,Schlickriede2018}.

We demonstrate how such nonlinear phase-matched metamaterials can be fabricated by stacking metasurfaces into three-dimensional (3D) structures, and show how the approach can considerably improve the performance of existing nonlinear metasurfaces. Our approach utilizes both local-field enhancement and the phase engineering of LSPRs. The latter provides more freedom to phase match nonlinear process than what is possible using conventional nonlinear materials (see Fig.~\ref{fig:Intro}b). We demonstrate both capabilities by fabricating metamaterial devices consisting of up to five layers of metasurfaces, that are phase-matched to emit SHG in the backward direction. The expected quadratic dependence of the emitted SHG signals on the number of stacked metasurfaces is demonstrated. 

For conventional materials and SHG, the phase changes are associated with the propagation of the fundamental and second-harmonic fields through the material.
Phase matching in such materials is connected to wavevector mismatch $\Delta k$ which vanishes for perfectly phase-matched processes. With $\Delta k$, we  can define phase-matching condition, e.g., for back-propagating SHG as $\Delta k=2k_{\omega}+k_{2\omega}=0$, where $k_{\omega}=n_\omega\omega/c$ and $k_{2\omega}=n_{2\omega}2\omega/c$ are the wavevector amplitudes at the fundamental ($\omega$) and SHG ($2\omega$) frequencies, respectively \cite{BoydBook2020}. With conventional nonlinear materials, this condition cannot be fulfilled, but it can be compensated by fabricating periodic quasi-phase-matched crystals~\cite{BoydBook2020, Pelton2004, Hu2013}. It is also possible to utilize zero-index materials to realize structures that have relaxed phase-matching requirements~\cite{Suchowski2013}.

\begin{figure}[htbp]
\centering
{\includegraphics[]{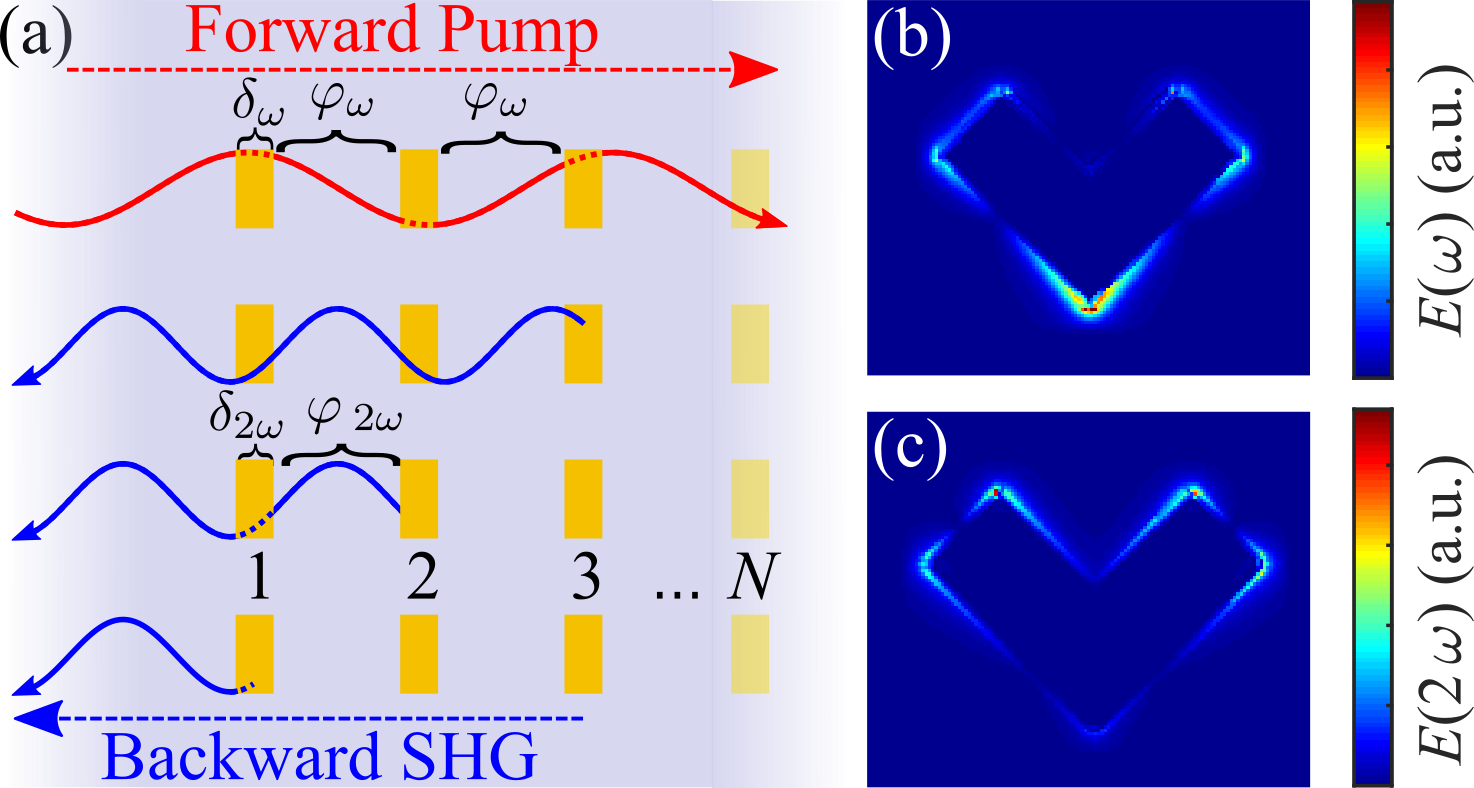}}
\caption{(a) Backward phase-matched SHG emission from  metamaterials consisting of $N$ stacked layers. The phase-matching condition is fulfilled by controlling the phase accumulation for both incident light (red arrows) and SHG light (blue arrows). The terms $\varphi_\omega$ and $\varphi_{2\omega}$ correspond to phase accumulation due to propagation.The LSPRs of metal nanoantennas enhance the local fields at (b) fundamental frequency $E_{\omega}$ and (c) SH frequency $E_{2\omega}$ and also induce the phase changes associated with LSPRs $\delta_\omega$ and $\delta_{2\omega}$, respectively.}
\label{fig:Sample}
\end{figure}

By using resonant metamaterials, we can extend this phase-matching condition by taking into account the phase changes $\delta_\omega$ and $\delta_{2\omega}$ that occur in a metamaterial due to coherent scattering of light from the constituent nanoantennas at the fundamental and SHG frequencies, respectively. Because these terms are dictated by the optical response of the nanoantennas, namely by their LSPRs~\cite{Genevet2017}, the extended phase-matching condition becomes solvable by metamaterial design. 
In order to demonstrate this capability, we designed and fabricated metamaterial devices where the backward SHG emission is phase-matched (see Fig.~\ref{fig:Sample}a).

%%%%%%%%%%%%%%%%%%%%
%%%%%%%%%%%%%%%%%%%%

The designed metamaterial devices consisted of a number of identical metasurfaces that were separated by identical spacer layers of thickness $h$. For such devices, the accumulated phase of the backward emitted SHG field should be a multiple of $2\pi$ resulting in the condition 
\begin{equation}
     \label{eq:PM-condition_reflection}
     2(\varphi_{\omega}+\delta_{\omega})+\varphi_{2\omega}+\delta_{2\omega}=2\pi m \,,
\end{equation}
where $m$ is an integer and terms $\varphi_{2\omega}=k_{2\omega}h$ and $\varphi_{\omega}=k_{\omega}h$ arise from the propagation of the fields. By now estimating the phase terms $\delta_\omega$ and $\delta_{2\omega}$ for the particles of interest, Eq.~(\ref{eq:PM-condition_reflection}) allows to solve for the spacer thickness $h$. The phase terms were numerically estimated by using the rigorous coupled wave analysis~\cite{Glytsis1987, Lalanne2000}. 

\begin{figure}[htbp]
\centering
\includegraphics[]{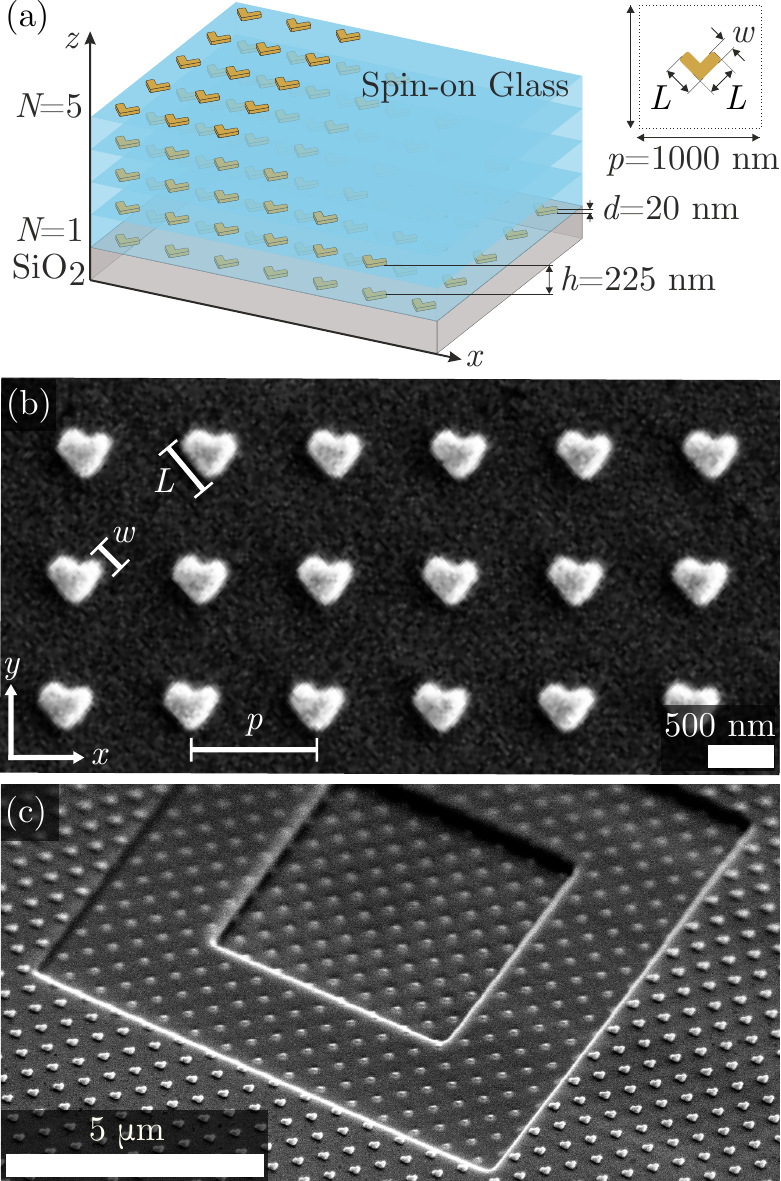}
\caption{(a) Investigated devices were composed of up to $N$ metasurfaces stacked on top of each other, separated by  $h=225$~nm thick spacer layers. Each of the metasurfaces consisted of a square array of 20 nm-thick V-shaped gold nanoantennas. (b) Representative scanning electron micrograph of one fabricated device (L180-3). (c) Oblique scanning electron micrograph obtained after successive etching with a focused ion beam, illustrating the stacked nature of the investigated metamaterial devices.}
\label{fig:FabricationSchematic}
\end{figure}

Our metamaterials consisted of a varying number $N$ metasurfaces composed of V-shaped gold nanoantennas with arm lengths of $L=180$~nm (L180-$N$) and $L=190$~nm (L190-$N$), arm widths of $w=100$~nm, and thicknesses of $d=20$~nm. These nanostructures were arranged into square lattices with a lattice constant of $p=1000$~nm (Fig.~\ref{fig:FabricationSchematic}a). This lattice configuration was chosen because it has been earlier found to emit SHG strongly~\cite{Czaplicki2018}. The above parameters were calculated to give rise to LSPRs centered near 1060~nm. According to Eq.~(\ref{eq:PM-condition_reflection}), for $m=0$ the phase-matching condition was fulfilled close to the LSPR wavelength by choosing the layer thickness of $h=225$~nm. Specifically, the phase-matching condition for devices L180-$N$ was fulfilled for linear input polarization orthogonal to the symmetry axis of the V-particles ($x$-axis) (Fig.~\ref{fig:FabricationSchematic}b). For devices L190-$N$ the condition was fulfilled for linear input polarization along the symmetry axis ($y$-axis) (Fig.~\ref{fig:FabricationSchematic}b).
Due to the symmetry properties of the samples, the generated SHG emission is polarized along the symmetry axis ($y$-axis) for all devices.

The devices were fabricated on a cleaned SiO$_2$ substrate through a sequence of steps repeated $N$ times~\cite{Yoon2017}. The sequence consists of the following eight steps: i) spin-coating polymethyl methacrylate (PMMA) resist EL8 at 5000 rpm speed for one minute followed by baking the sample on a hot plate at 150$^\circ$C for five minutes. ii) Spin-coating PMMA resist A2 at 2000 rpm for one minute and baking at 180$^\circ$C for five minutes, iii) spin-coating a third layer of conductive polymer (E-spacer) at 2000 rpm for one minute, in order to avoid charging effects during fabrication due to the insulating substrate. iv) Electron beam lithography of the nanostructures and bathing in deionized water in order to remove the E-spacer. v) Development in methyl isobutyl ketone:isopropanol (1:3) solution for 12 minutes at 0$^\circ$C followed by rinsing in isopropanol. vi) Deposition of a thin chromium layer (3 nm) and a layer of gold (20 nm) on the patterned resist by electron beam evaporation at 1 \AA/s. vii) Lift-off by bathing with acetone at 50$^\circ$C for one minute. viii) Spin-coating spacer layer (spin-on-glass IC1-200) at 6000 rpm for one minute followed by baking at 250$^\circ$C for five minutes in order to obtain a $h=225$~nm thick spacer layer. Representative scanning electron micrographs of one realized metamaterial device (L180-3) are shown in Fig.~\ref{fig:FabricationSchematic}b--c. 

%%%%%%%%%%%%%%%%%%%%
%%%%%%%%%%%%%%%%%%%%

\begin{figure}[htbp]
\centering
{\includegraphics[width=0.9\linewidth]{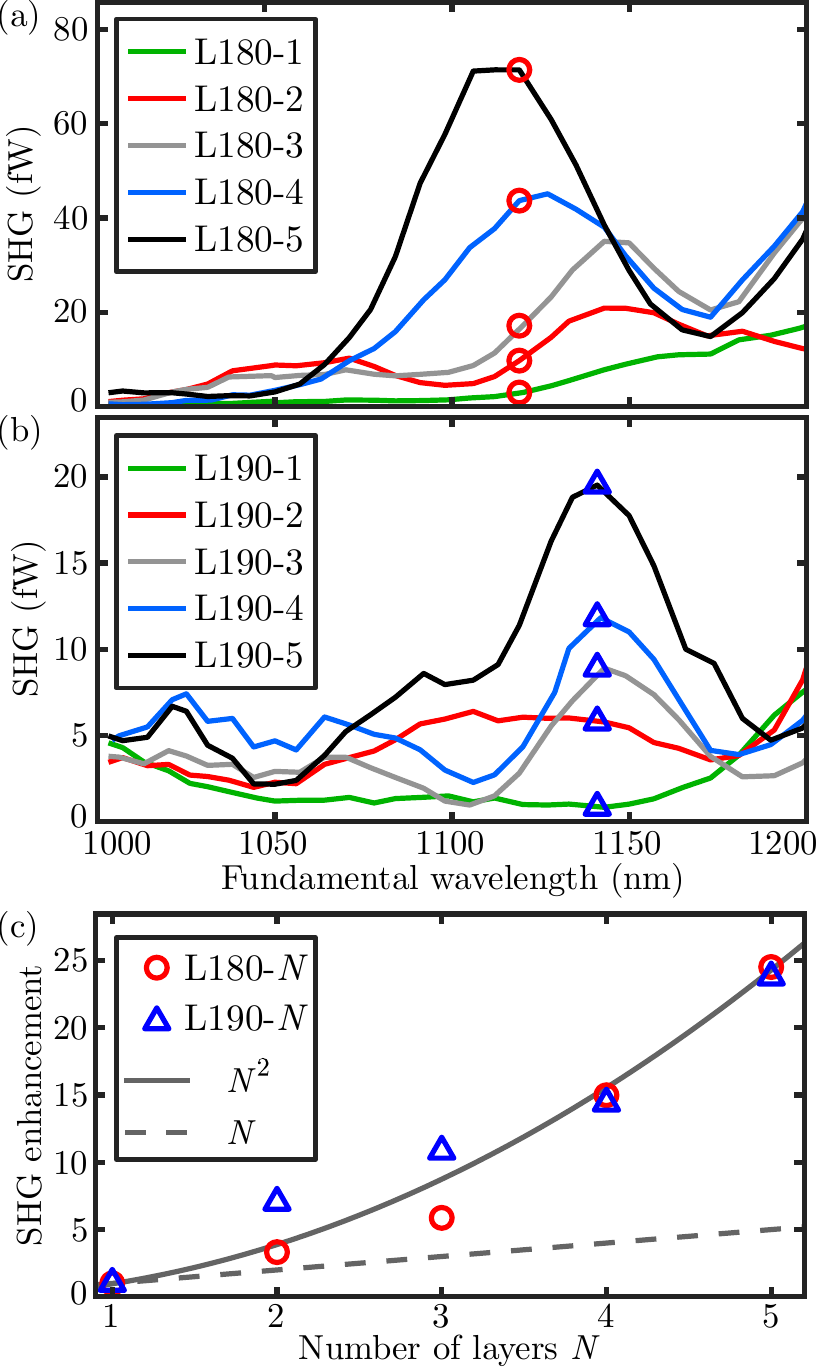}} 
\caption{Measured SHG emission power spectra from two sets of metamaterial devices (a) L180-$N$ and (b) L190-$N$. For devices L180-$N$ (L190-$N$), the constructive phase matching occurs near the pump wavelength of 1120~nm (1140~nm). (c) Calculated SHG enhancement at 1120~nm and 1140~nm for devices L180-$N$ (red circles) and L190-$N$ (blue triangles). Enhancements for devices L180-$N$ (L190-$N$) are normalized to the SHG signals detected from device L180-1 (L190-1). Enhancements are not proportional to $N$ (grey dashed line), but rather follow a quadratic trend (grey solid line).}
\label{fig:SHG}
\end{figure}

The SHG responses of the devices were characterized using a setup described in detail elsewhere~\cite{Czaplicki2018}. Briefly, a fs-laser oscillator (Chameleon Vision II, Ti:sapphire, 80~MHz) combined with an optical parametric oscillator (Chameleon Compact, 1000--1300~nm) was used as the pump, while the backward-emitted SHG signals were measured using a power-calibrated photomultiplier tube. See Supplemental Material at [URL will be inserted by publisher] for more detailed description of the setup. Here, we limited our input mean power to 10 mW in order to avoid possible sample damage. The SHG responses of the fabricated metamaterial devices (L180-$N$ and L190-$N$) consisting of varying number of metasurfaces ($N=$ 1, 2,..., 5) were measured as a function of the pump wavelength (see Fig.~\ref{fig:SHG}). 
%%%%%%%%%%%%%%%%%%%%
%%%%%%%%%%%%%%%%%%%%

The measured backward emitted SHG signals from the two sets of devices L180-$N$ (Fig.~\ref{fig:SHG}a) and L190-$N$ (Fig.~\ref{fig:SHG}b) both show a clear increase of the average SHG power when the number of metasurfaces ($N$) grows. The device with arm length of 180 nm composed of five metasurfaces (L180-5) resulted in the strongest signal corresponding to SHG power of 70 fW. When comparing the SHG responses from the two different sets of metamaterial devices (L180 and L190), one notices that the SHG enhancement as a function of $N$ is strongest (weakest) close to the wavelength range 1100--1150~nm (1000--1050~nm). We attribute these regions to be where the constructive (destructive) phase matching occurs. A closer analysis of the results reveals that the SHG responses at the fundamental wavelengths near 1120~nm (for L180-$N$) and 1140~nm (for L190-$N$) no longer depend linearly on the number of metasurfaces $N$ (Figs.~\ref{fig:SHG}a and~\ref{fig:SHG}b). Instead, the SHG signals follow close-to-quadratic dependence on $N$ (SHG $\propto N^{2}$) confirming that the devices were successfully phase matched (Fig.~\ref{fig:SHG}c). Furthermore, the devices were successfully phase matched in the challenging backward direction \cite{Liu2018}.

In addition to the quadratic dependence of the measured SHG signals on $N$, the SHG peaks of devices L180-$N$ blueshift from 1150 nm to 1120 nm when $N$ increases from three to five (Fig.~\ref{fig:SHG}a). Such behavior can be attributed to optical coupling of adjacent metasurfaces~\cite{Huttunen2016a, Segal2015}. It is known that adjacent particles in planar metasurfaces can become optically coupled forming collective responses known as surface lattice resonances, that have been found to enhance SHG emissions from metasurfaces~\cite{Michaeli2017b, Huttunen2018b, Czaplicki2018, Hooper2018}. It is only expected that similar effects could occur in 3D metamaterials also along the propagation direction~\cite{Segal2015}. In fact, it seems plausible that the detected very close to quadratic dependence of SHG signals on $N$ is also an outcome of this optical coupling mechanism, because the ideal quadratic dependence occurs only for materials with negligible losses. However, the fabricated devices exhibited losses, that were estimated by measuring the transmittance of a single metasurface to be close to 90\% near the pump wavelengths of 1120--1140~nm (See the measured transmittance spectra in the Supplemental Material at [URL will be inserted by publisher]). By taking into account this reduction in the pump intensity for subsequent metasurfaces, one would expect only around 12-fold SHG enhancement. Instead, we measured a 25-fold enhancement from both devices L180-$N$, and L190-$N$. The fact that the measured SHG enhancement was clearly above the simple estimation suggests that adjacent metasurfaces may have been already optically coupled.  

In this proof-of-principle demonstration, the relative positions of adjacent metasurfaces were not fully controlled. Therefore, we did not yet realize devices that would have allowed to more carefully investigate and utilize the radiative coupling mechanism between adjacent metasurfaces in order to further enhance the SHG emission. However, in the future it will be very interesting to investigate how to utilize this inter-metasurface optical coupling. For example, such coupling mechanism might allow designing nonlinear metamaterials where the SHG emission scales more favorably with the number of metasurfaces (SHG $\propto N^{n>2}$) than what is possible by using conventional nonlinear materials and their at most quadratic dependence ($n=2$) on the device length~\cite{BoydBook2020}.

In addition to enhancing the overall conversion efficiencies of nonlinear metamaterials, this demonstration of phase engineered nonlinear metamaterials has several other fundamental implications. For example, one can envisage how nonlinear metamaterials could be utilized for adiabatic frequency conversion, enabling broadband frequency conversion in nanomaterials~\cite{Suchowski2008,Suchowski2009}. Furthermore, this methodology could allow designing more efficient nonlinear terahertz-emitting metamaterials~\cite{Polyushkin2011, Luo2014, Keren-Zur2019a}.
Finally, the presented phase-engineering principles apply also for arbitrary wavefronts. Successful phase matching of nonlinear processes using complex spatial modes would have applications in holography and quantum computing~\cite{Li2015,Li2017}.

To conclude, we have demonstrated how the performance of nonlinear metamaterials can be substantially increased by stacking metasurfaces into three-dimensional metamaterials. Phase-matching considerations that are often difficult to fulfill using conventional materials can be easily solved by controlling the dimensions of the nanoantennas and the separation between the metasurfaces. We demonstrated this by phase matching second-harmonic generation emission from fabricated metamaterials in the challenging backward direction. We fabricated nonlinear metamaterial devices consisting of up to five stacked metasurfaces and demonstrated an up to 25-fold increase in the backward emitted second-harmonic intensities from the devices. Our results open a new paradigm of phase engineered three-dimensional nonlinear metamaterials, that could be used for example to realize more efficient nonlinear metamaterials.

\section*{Acknowledgements}
\noindent We acknowledge the support of the Academy of Finland (Grant No. 308596), the Flagship of Photonics Research and Innovation (PREIN) funded by the Academy of Finland (Grant No. 320165), the European Research Council (ERC) under the European Union's Horizon 2020 research and innovation programme (Grant agreements no. 639109 and no. 724881), and the National Research Foundation (NRF) grants (NRF-2019R1A2C3003129, NRF-2019M3A6B3030637, NRF-2019R1A5A8080290) funded by the Ministry of Science and ICT (MSIT) of the Republic of Korea.

\bibliography{NLOmetasurfaces}

\begin{thebibliography}{46}%
\makeatletter
\providecommand \@ifxundefined [1]{%
 \@ifx{#1\undefined}
}%
\providecommand \@ifnum [1]{%
 \ifnum #1\expandafter \@firstoftwo
 \else \expandafter \@secondoftwo
 \fi
}%
\providecommand \@ifx [1]{%
 \ifx #1\expandafter \@firstoftwo
 \else \expandafter \@secondoftwo
 \fi
}%
\providecommand \natexlab [1]{#1}%
\providecommand \enquote  [1]{``#1''}%
\providecommand \bibnamefont  [1]{#1}%
\providecommand \bibfnamefont [1]{#1}%
\providecommand \citenamefont [1]{#1}%
\providecommand \href@noop [0]{\@secondoftwo}%
\providecommand \href [0]{\begingroup \@sanitize@url \@href}%
\providecommand \@href[1]{\@@startlink{#1}\@@href}%
\providecommand \@@href[1]{\endgroup#1\@@endlink}%
\providecommand \@sanitize@url [0]{\catcode `\\12\catcode `\$12\catcode
  `\&12\catcode `\#12\catcode `\^12\catcode `\_12\catcode `\%12\relax}%
\providecommand \@@startlink[1]{}%
\providecommand \@@endlink[0]{}%
\providecommand \url  [0]{\begingroup\@sanitize@url \@url }%
\providecommand \@url [1]{\endgroup\@href {#1}{\urlprefix }}%
\providecommand \urlprefix  [0]{URL }%
\providecommand \Eprint [0]{\href }%
\providecommand \doibase [0]{http://dx.doi.org/}%
\providecommand \selectlanguage [0]{\@gobble}%
\providecommand \bibinfo  [0]{\@secondoftwo}%
\providecommand \bibfield  [0]{\@secondoftwo}%
\providecommand \translation [1]{[#1]}%
\providecommand \BibitemOpen [0]{}%
\providecommand \bibitemStop [0]{}%
\providecommand \bibitemNoStop [0]{.\EOS\space}%
\providecommand \EOS [0]{\spacefactor3000\relax}%
\providecommand \BibitemShut  [1]{\csname bibitem#1\endcsname}%
\let\auto@bib@innerbib\@empty
%</preamble>
\bibitem [{\citenamefont {Soukoulis}\ and\ \citenamefont
  {Wegener}(2011)}]{Soukoulis2011}%
  \BibitemOpen
  \bibfield  {author} {\bibinfo {author} {\bibfnamefont {C.~M.}\ \bibnamefont
  {Soukoulis}}\ and\ \bibinfo {author} {\bibfnamefont {M.}~\bibnamefont
  {Wegener}},\ }\href {http://dx.doi.org/10.1038/nphoton.2011.154
  http://10.0.4.14/nphoton.2011.154} {\bibfield  {journal} {\bibinfo  {journal}
  {Nat. Photonics}\ }\textbf {\bibinfo {volume} {5}},\ \bibinfo {pages} {523}
  (\bibinfo {year} {2011})}\BibitemShut {NoStop}%
\bibitem [{\citenamefont {Klein}\ \emph {et~al.}(2006)\citenamefont {Klein},
  \citenamefont {Enkrich}, \citenamefont {Wegener},\ and\ \citenamefont
  {Linden}}]{Klein2006}%
  \BibitemOpen
  \bibfield  {author} {\bibinfo {author} {\bibfnamefont {M.~W.}\ \bibnamefont
  {Klein}}, \bibinfo {author} {\bibfnamefont {C.}~\bibnamefont {Enkrich}},
  \bibinfo {author} {\bibfnamefont {M.}~\bibnamefont {Wegener}}, \ and\
  \bibinfo {author} {\bibfnamefont {S.}~\bibnamefont {Linden}},\ }\href@noop {}
  {\bibfield  {journal} {\bibinfo  {journal} {Science}\ }\textbf {\bibinfo
  {volume} {313}},\ \bibinfo {pages} {502} (\bibinfo {year}
  {2006})}\BibitemShut {NoStop}%
\bibitem [{\citenamefont {Al{\`{u}}}\ \emph {et~al.}(2007)\citenamefont
  {Al{\`{u}}}, \citenamefont {Silveirinha}, \citenamefont {Salandrino},\ and\
  \citenamefont {Engheta}}]{Alu2007}%
  \BibitemOpen
  \bibfield  {author} {\bibinfo {author} {\bibfnamefont {A.}~\bibnamefont
  {Al{\`{u}}}}, \bibinfo {author} {\bibfnamefont {M.~G.}\ \bibnamefont
  {Silveirinha}}, \bibinfo {author} {\bibfnamefont {A.}~\bibnamefont
  {Salandrino}}, \ and\ \bibinfo {author} {\bibfnamefont {N.}~\bibnamefont
  {Engheta}},\ }\href {\doibase 10.1103/PhysRevB.75.155410} {\bibfield
  {journal} {\bibinfo  {journal} {Phys. Rev. B}\ }\textbf {\bibinfo {volume}
  {75}},\ \bibinfo {pages} {155410} (\bibinfo {year} {2007})}\BibitemShut
  {NoStop}%
\bibitem [{\citenamefont {Zhang}\ \emph {et~al.}(2009)\citenamefont {Zhang},
  \citenamefont {Park}, \citenamefont {Li}, \citenamefont {Lu}, \citenamefont
  {Zhan},\ and\ \citenamefont {Zhang}}]{Zhang2009}%
  \BibitemOpen
  \bibfield  {author} {\bibinfo {author} {\bibfnamefont {S.}~\bibnamefont
  {Zhang}}, \bibinfo {author} {\bibfnamefont {Y.-S.}\ \bibnamefont {Park}},
  \bibinfo {author} {\bibfnamefont {J.}~\bibnamefont {Li}}, \bibinfo {author}
  {\bibfnamefont {X.}~\bibnamefont {Lu}}, \bibinfo {author} {\bibfnamefont
  {W.}~\bibnamefont {Zhan}}, \ and\ \bibinfo {author} {\bibfnamefont
  {X.}~\bibnamefont {Zhang}},\ }\href {\doibase 10.1109/PHO.2011.6110515}
  {\bibfield  {journal} {\bibinfo  {journal} {Phys. Rev. Lett.}\ }\textbf
  {\bibinfo {volume} {102}},\ \bibinfo {pages} {023901} (\bibinfo {year}
  {2009})}\BibitemShut {NoStop}%
\bibitem [{\citenamefont {Yu}\ and\ \citenamefont {Capasso}(2014)}]{Yu2014}%
  \BibitemOpen
  \bibfield  {author} {\bibinfo {author} {\bibfnamefont {N.}~\bibnamefont
  {Yu}}\ and\ \bibinfo {author} {\bibfnamefont {F.}~\bibnamefont {Capasso}},\
  }\href {\doibase 10.1038/nmat3839} {\bibfield  {journal} {\bibinfo  {journal}
  {Nat. Mater.}\ }\textbf {\bibinfo {volume} {13}},\ \bibinfo {pages} {139}
  (\bibinfo {year} {2014})}\BibitemShut {NoStop}%
\bibitem [{\citenamefont {Yin}\ \emph {et~al.}(2017)\citenamefont {Yin},
  \citenamefont {Steinle}, \citenamefont {Huang}, \citenamefont {Taubner},
  \citenamefont {Wuttig}, \citenamefont {Zentgraf},\ and\ \citenamefont
  {Giessen}}]{Yin2017}%
  \BibitemOpen
  \bibfield  {author} {\bibinfo {author} {\bibfnamefont {X.}~\bibnamefont
  {Yin}}, \bibinfo {author} {\bibfnamefont {T.}~\bibnamefont {Steinle}},
  \bibinfo {author} {\bibfnamefont {L.}~\bibnamefont {Huang}}, \bibinfo
  {author} {\bibfnamefont {T.}~\bibnamefont {Taubner}}, \bibinfo {author}
  {\bibfnamefont {M.}~\bibnamefont {Wuttig}}, \bibinfo {author} {\bibfnamefont
  {T.}~\bibnamefont {Zentgraf}}, \ and\ \bibinfo {author} {\bibfnamefont
  {H.}~\bibnamefont {Giessen}},\ }\href {\doibase 10.1038/lsa.2017.16}
  {\bibfield  {journal} {\bibinfo  {journal} {Light Sci. Appl.}\ }\textbf
  {\bibinfo {volume} {6}},\ \bibinfo {pages} {e17016} (\bibinfo {year}
  {2017})}\BibitemShut {NoStop}%
\bibitem [{\citenamefont {Genevet}\ \emph {et~al.}(2017)\citenamefont
  {Genevet}, \citenamefont {Capasso}, \citenamefont {Aieta}, \citenamefont
  {Khorasaninejad},\ and\ \citenamefont {Devlin}}]{Genevet2017}%
  \BibitemOpen
  \bibfield  {author} {\bibinfo {author} {\bibfnamefont {P.}~\bibnamefont
  {Genevet}}, \bibinfo {author} {\bibfnamefont {F.}~\bibnamefont {Capasso}},
  \bibinfo {author} {\bibfnamefont {F.}~\bibnamefont {Aieta}}, \bibinfo
  {author} {\bibfnamefont {M.}~\bibnamefont {Khorasaninejad}}, \ and\ \bibinfo
  {author} {\bibfnamefont {R.}~\bibnamefont {Devlin}},\ }\href {\doibase
  10.1364/OPTICA.4.000139} {\bibfield  {journal} {\bibinfo  {journal} {Optica}\
  }\textbf {\bibinfo {volume} {4}},\ \bibinfo {pages} {139} (\bibinfo {year}
  {2017})}\BibitemShut {NoStop}%
\bibitem [{\citenamefont {Huang}\ \emph {et~al.}(2018)\citenamefont {Huang},
  \citenamefont {Zhang},\ and\ \citenamefont {Zentgraf}}]{Huang2018}%
  \BibitemOpen
  \bibfield  {author} {\bibinfo {author} {\bibfnamefont {L.}~\bibnamefont
  {Huang}}, \bibinfo {author} {\bibfnamefont {S.}~\bibnamefont {Zhang}}, \ and\
  \bibinfo {author} {\bibfnamefont {T.}~\bibnamefont {Zentgraf}},\ }\href
  {\doibase 10.1515/nanoph-2017-0118} {\bibfield  {journal} {\bibinfo
  {journal} {Nanophotonics}\ }\textbf {\bibinfo {volume} {7}},\ \bibinfo
  {pages} {1169} (\bibinfo {year} {2018})}\BibitemShut {NoStop}%
\bibitem [{\citenamefont {Arbabi}\ \emph {et~al.}(2018)\citenamefont {Arbabi},
  \citenamefont {Kamali}, \citenamefont {Arbabi},\ and\ \citenamefont
  {Faraon}}]{Arbabi2018}%
  \BibitemOpen
  \bibfield  {author} {\bibinfo {author} {\bibfnamefont {E.}~\bibnamefont
  {Arbabi}}, \bibinfo {author} {\bibfnamefont {S.~M.}\ \bibnamefont {Kamali}},
  \bibinfo {author} {\bibfnamefont {A.}~\bibnamefont {Arbabi}}, \ and\ \bibinfo
  {author} {\bibfnamefont {A.}~\bibnamefont {Faraon}},\ }\href {\doibase
  10.1021/acsphotonics.8b00362} {\bibfield  {journal} {\bibinfo  {journal} {ACS
  Photonics}\ }\textbf {\bibinfo {volume} {5}},\ \bibinfo {pages} {3132}
  (\bibinfo {year} {2018})}\BibitemShut {NoStop}%
\bibitem [{\citenamefont {Ren}\ \emph {et~al.}(2019)\citenamefont {Ren},
  \citenamefont {Briere}, \citenamefont {Fang}, \citenamefont {Ni},
  \citenamefont {Sawant}, \citenamefont {H{\'{e}}ron}, \citenamefont {Chenot},
  \citenamefont {V{\'{e}}zian}, \citenamefont {Damilano}, \citenamefont
  {Br{\"{a}}ndli}, \citenamefont {Maier},\ and\ \citenamefont
  {Genevet}}]{Ren2019}%
  \BibitemOpen
  \bibfield  {author} {\bibinfo {author} {\bibfnamefont {H.}~\bibnamefont
  {Ren}}, \bibinfo {author} {\bibfnamefont {G.}~\bibnamefont {Briere}},
  \bibinfo {author} {\bibfnamefont {X.}~\bibnamefont {Fang}}, \bibinfo {author}
  {\bibfnamefont {P.}~\bibnamefont {Ni}}, \bibinfo {author} {\bibfnamefont
  {R.}~\bibnamefont {Sawant}}, \bibinfo {author} {\bibfnamefont
  {S.}~\bibnamefont {H{\'{e}}ron}}, \bibinfo {author} {\bibfnamefont
  {S.}~\bibnamefont {Chenot}}, \bibinfo {author} {\bibfnamefont
  {S.}~\bibnamefont {V{\'{e}}zian}}, \bibinfo {author} {\bibfnamefont
  {B.}~\bibnamefont {Damilano}}, \bibinfo {author} {\bibfnamefont
  {V.}~\bibnamefont {Br{\"{a}}ndli}}, \bibinfo {author} {\bibfnamefont {S.~A.}\
  \bibnamefont {Maier}}, \ and\ \bibinfo {author} {\bibfnamefont
  {P.}~\bibnamefont {Genevet}},\ }\href {\doibase 10.1038/s41467-019-11030-1}
  {\bibfield  {journal} {\bibinfo  {journal} {Nat. Commun.}\ }\textbf {\bibinfo
  {volume} {10}},\ \bibinfo {pages} {2986} (\bibinfo {year}
  {2019})}\BibitemShut {NoStop}%
\bibitem [{\citenamefont {Rubin}\ \emph {et~al.}(2019)\citenamefont {Rubin},
  \citenamefont {D'Aversa}, \citenamefont {Chevalier}, \citenamefont {Shi},
  \citenamefont {Chen},\ and\ \citenamefont {Capasso}}]{Rubin2019}%
  \BibitemOpen
  \bibfield  {author} {\bibinfo {author} {\bibfnamefont {N.~A.}\ \bibnamefont
  {Rubin}}, \bibinfo {author} {\bibfnamefont {G.}~\bibnamefont {D'Aversa}},
  \bibinfo {author} {\bibfnamefont {P.}~\bibnamefont {Chevalier}}, \bibinfo
  {author} {\bibfnamefont {Z.}~\bibnamefont {Shi}}, \bibinfo {author}
  {\bibfnamefont {W.~T.}\ \bibnamefont {Chen}}, \ and\ \bibinfo {author}
  {\bibfnamefont {F.}~\bibnamefont {Capasso}},\ }\href {\doibase
  10.1126/science.aax1839} {\bibfield  {journal} {\bibinfo  {journal}
  {Science}\ }\textbf {\bibinfo {volume} {365}},\ \bibinfo {pages} {43}
  (\bibinfo {year} {2019})}\BibitemShut {NoStop}%
\bibitem [{\citenamefont {Kwiat}\ \emph {et~al.}(1995)\citenamefont {Kwiat},
  \citenamefont {Mattle}, \citenamefont {Weinfurter}, \citenamefont
  {Zeilinger}, \citenamefont {Sergienko},\ and\ \citenamefont
  {Shih}}]{Kwiat1995}%
  \BibitemOpen
  \bibfield  {author} {\bibinfo {author} {\bibfnamefont {P.~G.}\ \bibnamefont
  {Kwiat}}, \bibinfo {author} {\bibfnamefont {K.}~\bibnamefont {Mattle}},
  \bibinfo {author} {\bibfnamefont {H.}~\bibnamefont {Weinfurter}}, \bibinfo
  {author} {\bibfnamefont {A.}~\bibnamefont {Zeilinger}}, \bibinfo {author}
  {\bibfnamefont {A.~V.}\ \bibnamefont {Sergienko}}, \ and\ \bibinfo {author}
  {\bibfnamefont {Y.}~\bibnamefont {Shih}},\ }\href {\doibase
  10.1103/PhysRevLett.75.4337} {\bibfield  {journal} {\bibinfo  {journal}
  {Phys. Rev. Lett.}\ }\textbf {\bibinfo {volume} {75}},\ \bibinfo {pages}
  {4337} (\bibinfo {year} {1995})}\BibitemShut {NoStop}%
\bibitem [{\citenamefont {Brabec}\ and\ \citenamefont
  {Krausz}(2000)}]{Brabec2000intense}%
  \BibitemOpen
  \bibfield  {author} {\bibinfo {author} {\bibfnamefont {T.}~\bibnamefont
  {Brabec}}\ and\ \bibinfo {author} {\bibfnamefont {F.}~\bibnamefont
  {Krausz}},\ }\href@noop {} {\bibfield  {journal} {\bibinfo  {journal} {Rev.
  Mod. Phys.}\ }\textbf {\bibinfo {volume} {72}},\ \bibinfo {pages} {545}
  (\bibinfo {year} {2000})}\BibitemShut {NoStop}%
\bibitem [{\citenamefont {Kippenberg}\ \emph {et~al.}(2011)\citenamefont
  {Kippenberg}, \citenamefont {Holzwarth},\ and\ \citenamefont
  {Diddams}}]{Kippenberg2011comb}%
  \BibitemOpen
  \bibfield  {author} {\bibinfo {author} {\bibfnamefont {T.~J.}\ \bibnamefont
  {Kippenberg}}, \bibinfo {author} {\bibfnamefont {R.}~\bibnamefont
  {Holzwarth}}, \ and\ \bibinfo {author} {\bibfnamefont {S.~A.}\ \bibnamefont
  {Diddams}},\ }\href@noop {} {\bibfield  {journal} {\bibinfo  {journal}
  {Science}\ }\textbf {\bibinfo {volume} {332}},\ \bibinfo {pages} {555}
  (\bibinfo {year} {2011})}\BibitemShut {NoStop}%
\bibitem [{\citenamefont {Shcherbakov}\ \emph {et~al.}(2015)\citenamefont
  {Shcherbakov}, \citenamefont {Vabishchevich}, \citenamefont {Shorokhov},
  \citenamefont {Chong}, \citenamefont {Choi}, \citenamefont {Staude},
  \citenamefont {Miroshnichenko}, \citenamefont {Neshev}, \citenamefont
  {Fedyanin},\ and\ \citenamefont {Kivshar}}]{Shcherbakov2015}%
  \BibitemOpen
  \bibfield  {author} {\bibinfo {author} {\bibfnamefont {M.~R.}\ \bibnamefont
  {Shcherbakov}}, \bibinfo {author} {\bibfnamefont {P.~P.}\ \bibnamefont
  {Vabishchevich}}, \bibinfo {author} {\bibfnamefont {A.~S.}\ \bibnamefont
  {Shorokhov}}, \bibinfo {author} {\bibfnamefont {K.~E.}\ \bibnamefont
  {Chong}}, \bibinfo {author} {\bibfnamefont {D.~Y.}\ \bibnamefont {Choi}},
  \bibinfo {author} {\bibfnamefont {I.}~\bibnamefont {Staude}}, \bibinfo
  {author} {\bibfnamefont {A.~E.}\ \bibnamefont {Miroshnichenko}}, \bibinfo
  {author} {\bibfnamefont {D.~N.}\ \bibnamefont {Neshev}}, \bibinfo {author}
  {\bibfnamefont {A.~A.}\ \bibnamefont {Fedyanin}}, \ and\ \bibinfo {author}
  {\bibfnamefont {Y.~S.}\ \bibnamefont {Kivshar}},\ }\href {\doibase
  10.1021/acs.nanolett.5b02989} {\bibfield  {journal} {\bibinfo  {journal}
  {Nano Lett.}\ }\textbf {\bibinfo {volume} {15}},\ \bibinfo {pages} {6985}
  (\bibinfo {year} {2015})}\BibitemShut {NoStop}%
\bibitem [{\citenamefont {Boyd}(2020)}]{BoydBook2020}%
  \BibitemOpen
  \bibfield  {author} {\bibinfo {author} {\bibfnamefont {R.~W.}\ \bibnamefont
  {Boyd}},\ }\href@noop {} {\emph {\bibinfo {title} {{Nonlinear optics}}}}\
  (\bibinfo  {publisher} {Academic Press},\ \bibinfo {address} {San Diego},\
  \bibinfo {year} {2020})\BibitemShut {NoStop}%
\bibitem [{\citenamefont {Wang}\ \emph {et~al.}(2017)\citenamefont {Wang},
  \citenamefont {Li}, \citenamefont {Kim}, \citenamefont {Xiong}, \citenamefont
  {Ren}, \citenamefont {Guo}, \citenamefont {Yu},\ and\ \citenamefont
  {Lon{\v{c}}ar}}]{Wang2017}%
  \BibitemOpen
  \bibfield  {author} {\bibinfo {author} {\bibfnamefont {C.}~\bibnamefont
  {Wang}}, \bibinfo {author} {\bibfnamefont {Z.}~\bibnamefont {Li}}, \bibinfo
  {author} {\bibfnamefont {M.~H.}\ \bibnamefont {Kim}}, \bibinfo {author}
  {\bibfnamefont {X.}~\bibnamefont {Xiong}}, \bibinfo {author} {\bibfnamefont
  {X.~F.}\ \bibnamefont {Ren}}, \bibinfo {author} {\bibfnamefont {G.~C.}\
  \bibnamefont {Guo}}, \bibinfo {author} {\bibfnamefont {N.}~\bibnamefont
  {Yu}}, \ and\ \bibinfo {author} {\bibfnamefont {M.}~\bibnamefont
  {Lon{\v{c}}ar}},\ }\href {\doibase 10.1038/s41467-017-02189-6} {\bibfield
  {journal} {\bibinfo  {journal} {Nat. Commun.}\ }\textbf {\bibinfo {volume}
  {8}},\ \bibinfo {pages} {2098} (\bibinfo {year} {2017})}\BibitemShut
  {NoStop}%
\bibitem [{\citenamefont {{Lim}}\ \emph {et~al.}(1989)\citenamefont {{Lim}},
  \citenamefont {{Fejer}},\ and\ \citenamefont {{Byer}}}]{Lim1989}%
  \BibitemOpen
  \bibfield  {author} {\bibinfo {author} {\bibfnamefont {E.~J.}\ \bibnamefont
  {{Lim}}}, \bibinfo {author} {\bibfnamefont {M.~M.}\ \bibnamefont {{Fejer}}},
  \ and\ \bibinfo {author} {\bibfnamefont {R.~L.}\ \bibnamefont {{Byer}}},\
  }\href@noop {} {\bibfield  {journal} {\bibinfo  {journal} {Electron. Lett.}\
  }\textbf {\bibinfo {volume} {25}},\ \bibinfo {pages} {174} (\bibinfo {year}
  {1989})}\BibitemShut {NoStop}%
\bibitem [{\citenamefont {Kauranen}\ and\ \citenamefont
  {Zayats}(2012)}]{Kauranen2012review}%
  \BibitemOpen
  \bibfield  {author} {\bibinfo {author} {\bibfnamefont {M.}~\bibnamefont
  {Kauranen}}\ and\ \bibinfo {author} {\bibfnamefont {A.~V.}\ \bibnamefont
  {Zayats}},\ }\href {\doibase 10.1016/B978-0-444-59526-3.00011-2} {\bibfield
  {journal} {\bibinfo  {journal} {Nat. Photonics}\ }\textbf {\bibinfo {volume}
  {6}},\ \bibinfo {pages} {737} (\bibinfo {year} {2012})}\BibitemShut {NoStop}%
\bibitem [{\citenamefont {Lee}\ \emph {et~al.}(2014)\citenamefont {Lee},
  \citenamefont {Tymchenko}, \citenamefont {Argyropoulos}, \citenamefont
  {Chen}, \citenamefont {Lu}, \citenamefont {Demmerle}, \citenamefont {Boehm},
  \citenamefont {Amann}, \citenamefont {Al{\`{u}}},\ and\ \citenamefont
  {Belkin}}]{Lee2014}%
  \BibitemOpen
  \bibfield  {author} {\bibinfo {author} {\bibfnamefont {J.}~\bibnamefont
  {Lee}}, \bibinfo {author} {\bibfnamefont {M.}~\bibnamefont {Tymchenko}},
  \bibinfo {author} {\bibfnamefont {C.}~\bibnamefont {Argyropoulos}}, \bibinfo
  {author} {\bibfnamefont {P.~Y.}\ \bibnamefont {Chen}}, \bibinfo {author}
  {\bibfnamefont {F.}~\bibnamefont {Lu}}, \bibinfo {author} {\bibfnamefont
  {F.}~\bibnamefont {Demmerle}}, \bibinfo {author} {\bibfnamefont
  {G.}~\bibnamefont {Boehm}}, \bibinfo {author} {\bibfnamefont {M.~C.}\
  \bibnamefont {Amann}}, \bibinfo {author} {\bibfnamefont {A.}~\bibnamefont
  {Al{\`{u}}}}, \ and\ \bibinfo {author} {\bibfnamefont {M.~A.}\ \bibnamefont
  {Belkin}},\ }\href {\doibase 10.1038/nature13455} {\bibfield  {journal}
  {\bibinfo  {journal} {Nature}\ }\textbf {\bibinfo {volume} {511}},\ \bibinfo
  {pages} {65} (\bibinfo {year} {2014})}\BibitemShut {NoStop}%
\bibitem [{\citenamefont {Maier}(2007)}]{Maier2007}%
  \BibitemOpen
  \bibfield  {author} {\bibinfo {author} {\bibfnamefont {S.~A.}\ \bibnamefont
  {Maier}},\ }\href@noop {} {\emph {\bibinfo {title} {{Plasmonics: fundamentals
  and applications}}}}\ (\bibinfo  {publisher} {Springer Science {\&} Business
  Media},\ \bibinfo {year} {2007})\BibitemShut {NoStop}%
\bibitem [{\citenamefont {Lapine}\ \emph {et~al.}(2014)\citenamefont {Lapine},
  \citenamefont {Shadrivov},\ and\ \citenamefont {Kivshar}}]{Lapine2014}%
  \BibitemOpen
  \bibfield  {author} {\bibinfo {author} {\bibfnamefont {M.}~\bibnamefont
  {Lapine}}, \bibinfo {author} {\bibfnamefont {I.~V.}\ \bibnamefont
  {Shadrivov}}, \ and\ \bibinfo {author} {\bibfnamefont {Y.~S.}\ \bibnamefont
  {Kivshar}},\ }\href {\doibase 10.1103/RevModPhys.86.1093} {\bibfield
  {journal} {\bibinfo  {journal} {Rev. Mod. Phys.}\ }\textbf {\bibinfo {volume}
  {86}},\ \bibinfo {pages} {1093} (\bibinfo {year} {2014})}\BibitemShut
  {NoStop}%
\bibitem [{\citenamefont {Butet}\ \emph {et~al.}(2015)\citenamefont {Butet},
  \citenamefont {Brevet},\ and\ \citenamefont {Martin}}]{ButetReview2015}%
  \BibitemOpen
  \bibfield  {author} {\bibinfo {author} {\bibfnamefont {J.}~\bibnamefont
  {Butet}}, \bibinfo {author} {\bibfnamefont {P.~F.}\ \bibnamefont {Brevet}}, \
  and\ \bibinfo {author} {\bibfnamefont {O.~J.}\ \bibnamefont {Martin}},\
  }\href {\doibase 10.1021/acsnano.5b04373} {\bibfield  {journal} {\bibinfo
  {journal} {ACS Nano}\ }\textbf {\bibinfo {volume} {9}},\ \bibinfo {pages}
  {10545} (\bibinfo {year} {2015})}\BibitemShut {NoStop}%
\bibitem [{\citenamefont {Li}\ \emph {et~al.}(2017)\citenamefont {Li},
  \citenamefont {Zhang},\ and\ \citenamefont {Zentgraf}}]{Li2017}%
  \BibitemOpen
  \bibfield  {author} {\bibinfo {author} {\bibfnamefont {G.}~\bibnamefont
  {Li}}, \bibinfo {author} {\bibfnamefont {S.}~\bibnamefont {Zhang}}, \ and\
  \bibinfo {author} {\bibfnamefont {T.}~\bibnamefont {Zentgraf}},\ }\href
  {\doibase 10.1038/natrevmats.2017.10} {\bibfield  {journal} {\bibinfo
  {journal} {Nat. Rev. Mater.}\ }\textbf {\bibinfo {volume} {2}},\ \bibinfo
  {pages} {1} (\bibinfo {year} {2017})}\BibitemShut {NoStop}%
\bibitem [{\citenamefont {Rahimi}\ and\ \citenamefont
  {Gordon}(2018)}]{Rahimi2018}%
  \BibitemOpen
  \bibfield  {author} {\bibinfo {author} {\bibfnamefont {E.}~\bibnamefont
  {Rahimi}}\ and\ \bibinfo {author} {\bibfnamefont {R.}~\bibnamefont
  {Gordon}},\ }\href {\doibase 10.1002/adom.201800274} {\bibfield  {journal}
  {\bibinfo  {journal} {Adv. Opt. Mater.}\ }\textbf {\bibinfo {volume} {6}},\
  \bibinfo {pages} {1} (\bibinfo {year} {2018})}\BibitemShut {NoStop}%
\bibitem [{\citenamefont {Huttunen}\ \emph {et~al.}(2019)\citenamefont
  {Huttunen}, \citenamefont {Czaplicki},\ and\ \citenamefont
  {Kauranen}}]{Huttunen2019review}%
  \BibitemOpen
  \bibfield  {author} {\bibinfo {author} {\bibfnamefont {M.~J.}\ \bibnamefont
  {Huttunen}}, \bibinfo {author} {\bibfnamefont {R.}~\bibnamefont {Czaplicki}},
  \ and\ \bibinfo {author} {\bibfnamefont {M.}~\bibnamefont {Kauranen}},\
  }\href {\doibase 10.1142/S0218863519500012} {\bibfield  {journal} {\bibinfo
  {journal} {J. Nonlinear Opt. Phys. Mater.}\ }\textbf {\bibinfo {volume}
  {28}},\ \bibinfo {pages} {1950001} (\bibinfo {year} {2019})}\BibitemShut
  {NoStop}%
\bibitem [{\citenamefont {Li}\ \emph {et~al.}(2015)\citenamefont {Li},
  \citenamefont {Chen}, \citenamefont {Pholchai}, \citenamefont {Reineke},
  \citenamefont {Wong}, \citenamefont {Pun}, \citenamefont {Cheah},
  \citenamefont {Zentgraf},\ and\ \citenamefont {Zhang}}]{Li2015}%
  \BibitemOpen
  \bibfield  {author} {\bibinfo {author} {\bibfnamefont {G.}~\bibnamefont
  {Li}}, \bibinfo {author} {\bibfnamefont {S.}~\bibnamefont {Chen}}, \bibinfo
  {author} {\bibfnamefont {N.}~\bibnamefont {Pholchai}}, \bibinfo {author}
  {\bibfnamefont {B.}~\bibnamefont {Reineke}}, \bibinfo {author} {\bibfnamefont
  {P.~W.~H.}\ \bibnamefont {Wong}}, \bibinfo {author} {\bibfnamefont
  {E.~Y.~B.}\ \bibnamefont {Pun}}, \bibinfo {author} {\bibfnamefont {K.~W.}\
  \bibnamefont {Cheah}}, \bibinfo {author} {\bibfnamefont {T.}~\bibnamefont
  {Zentgraf}}, \ and\ \bibinfo {author} {\bibfnamefont {S.}~\bibnamefont
  {Zhang}},\ }\href {\doibase 10.1038/nmat4267} {\bibfield  {journal} {\bibinfo
   {journal} {Nat. Mater.}\ }\textbf {\bibinfo {volume} {14}},\ \bibinfo
  {pages} {607} (\bibinfo {year} {2015})}\BibitemShut {NoStop}%
\bibitem [{\citenamefont {Schlickriede}\ \emph {et~al.}(2018)\citenamefont
  {Schlickriede}, \citenamefont {Waterman}, \citenamefont {Reineke},
  \citenamefont {Georgi}, \citenamefont {Li}, \citenamefont {Zhang},\ and\
  \citenamefont {Zentgraf}}]{Schlickriede2018}%
  \BibitemOpen
  \bibfield  {author} {\bibinfo {author} {\bibfnamefont {C.}~\bibnamefont
  {Schlickriede}}, \bibinfo {author} {\bibfnamefont {N.}~\bibnamefont
  {Waterman}}, \bibinfo {author} {\bibfnamefont {B.}~\bibnamefont {Reineke}},
  \bibinfo {author} {\bibfnamefont {P.}~\bibnamefont {Georgi}}, \bibinfo
  {author} {\bibfnamefont {G.}~\bibnamefont {Li}}, \bibinfo {author}
  {\bibfnamefont {S.}~\bibnamefont {Zhang}}, \ and\ \bibinfo {author}
  {\bibfnamefont {T.}~\bibnamefont {Zentgraf}},\ }\href {\doibase
  10.1002/adma.201703843} {\bibfield  {journal} {\bibinfo  {journal} {Adv.
  Mater.}\ }\textbf {\bibinfo {volume} {30}},\ \bibinfo {pages} {1703843}
  (\bibinfo {year} {2018})}\BibitemShut {NoStop}%
\bibitem [{\citenamefont {Pelton}\ \emph {et~al.}(2004)\citenamefont {Pelton},
  \citenamefont {Marsden}, \citenamefont {Ljunggren}, \citenamefont {Tengner},
  \citenamefont {Karlsson}, \citenamefont {Fragemann}, \citenamefont
  {Canalias},\ and\ \citenamefont {Laurell}}]{Pelton2004}%
  \BibitemOpen
  \bibfield  {author} {\bibinfo {author} {\bibfnamefont {M.}~\bibnamefont
  {Pelton}}, \bibinfo {author} {\bibfnamefont {P.}~\bibnamefont {Marsden}},
  \bibinfo {author} {\bibfnamefont {D.}~\bibnamefont {Ljunggren}}, \bibinfo
  {author} {\bibfnamefont {M.}~\bibnamefont {Tengner}}, \bibinfo {author}
  {\bibfnamefont {A.}~\bibnamefont {Karlsson}}, \bibinfo {author}
  {\bibfnamefont {A.}~\bibnamefont {Fragemann}}, \bibinfo {author}
  {\bibfnamefont {C.}~\bibnamefont {Canalias}}, \ and\ \bibinfo {author}
  {\bibfnamefont {F.}~\bibnamefont {Laurell}},\ }\href {\doibase
  10.1364/OPEX.12.003573} {\bibfield  {journal} {\bibinfo  {journal} {Opt.
  Express}\ }\textbf {\bibinfo {volume} {12}},\ \bibinfo {pages} {3573}
  (\bibinfo {year} {2004})}\BibitemShut {NoStop}%
\bibitem [{\citenamefont {Hu}\ \emph {et~al.}(2013)\citenamefont {Hu},
  \citenamefont {Xu},\ and\ \citenamefont {Zhu}}]{Hu2013}%
  \BibitemOpen
  \bibfield  {author} {\bibinfo {author} {\bibfnamefont {X.~P.}\ \bibnamefont
  {Hu}}, \bibinfo {author} {\bibfnamefont {P.}~\bibnamefont {Xu}}, \ and\
  \bibinfo {author} {\bibfnamefont {S.~N.}\ \bibnamefont {Zhu}},\ }\href
  {\doibase 10.1364/PRJ.1.000171} {\bibfield  {journal} {\bibinfo  {journal}
  {Photonics Res.}\ }\textbf {\bibinfo {volume} {1}},\ \bibinfo {pages} {171}
  (\bibinfo {year} {2013})}\BibitemShut {NoStop}%
\bibitem [{\citenamefont {Suchowski}\ \emph {et~al.}(2013)\citenamefont
  {Suchowski}, \citenamefont {O'Brien}, \citenamefont {Wong}, \citenamefont
  {Salandrino}, \citenamefont {Yin},\ and\ \citenamefont
  {Zhang}}]{Suchowski2013}%
  \BibitemOpen
  \bibfield  {author} {\bibinfo {author} {\bibfnamefont {H.}~\bibnamefont
  {Suchowski}}, \bibinfo {author} {\bibfnamefont {K.}~\bibnamefont {O'Brien}},
  \bibinfo {author} {\bibfnamefont {Z.~J.}\ \bibnamefont {Wong}}, \bibinfo
  {author} {\bibfnamefont {A.}~\bibnamefont {Salandrino}}, \bibinfo {author}
  {\bibfnamefont {X.}~\bibnamefont {Yin}}, \ and\ \bibinfo {author}
  {\bibfnamefont {X.}~\bibnamefont {Zhang}},\ }\href@noop {} {\bibfield
  {journal} {\bibinfo  {journal} {Science}\ }\textbf {\bibinfo {volume}
  {342}},\ \bibinfo {pages} {1223} (\bibinfo {year} {2013})}\BibitemShut
  {NoStop}%
\bibitem [{\citenamefont {Glytsis}\ and\ \citenamefont
  {Gaylord}(1987)}]{Glytsis1987}%
  \BibitemOpen
  \bibfield  {author} {\bibinfo {author} {\bibfnamefont {E.~N.}\ \bibnamefont
  {Glytsis}}\ and\ \bibinfo {author} {\bibfnamefont {T.~K.}\ \bibnamefont
  {Gaylord}},\ }\href {\doibase 10.1364/josaa.4.002061} {\bibfield  {journal}
  {\bibinfo  {journal} {J. Opt. Soc. Am. A}\ }\textbf {\bibinfo {volume} {4}},\
  \bibinfo {pages} {2061} (\bibinfo {year} {1987})}\BibitemShut {NoStop}%
\bibitem [{\citenamefont {Lalanne}\ and\ \citenamefont
  {Silberstein}(2000)}]{Lalanne2000}%
  \BibitemOpen
  \bibfield  {author} {\bibinfo {author} {\bibfnamefont {P.}~\bibnamefont
  {Lalanne}}\ and\ \bibinfo {author} {\bibfnamefont {E.}~\bibnamefont
  {Silberstein}},\ }\href {\doibase 10.1364/OL.25.001092} {\bibfield  {journal}
  {\bibinfo  {journal} {Opt. Lett.}\ }\textbf {\bibinfo {volume} {25}},\
  \bibinfo {pages} {1092} (\bibinfo {year} {2000})}\BibitemShut {NoStop}%
\bibitem [{\citenamefont {Czaplicki}\ \emph {et~al.}(2018)\citenamefont
  {Czaplicki}, \citenamefont {Kiviniemi}, \citenamefont {Huttunen},
  \citenamefont {Zang}, \citenamefont {Stolt}, \citenamefont {Vartiainen},
  \citenamefont {Butet}, \citenamefont {Kuittinen}, \citenamefont {Martin},\
  and\ \citenamefont {Kauranen}}]{Czaplicki2018}%
  \BibitemOpen
  \bibfield  {author} {\bibinfo {author} {\bibfnamefont {R.}~\bibnamefont
  {Czaplicki}}, \bibinfo {author} {\bibfnamefont {A.}~\bibnamefont
  {Kiviniemi}}, \bibinfo {author} {\bibfnamefont {M.~J.}\ \bibnamefont
  {Huttunen}}, \bibinfo {author} {\bibfnamefont {X.}~\bibnamefont {Zang}},
  \bibinfo {author} {\bibfnamefont {T.}~\bibnamefont {Stolt}}, \bibinfo
  {author} {\bibfnamefont {I.}~\bibnamefont {Vartiainen}}, \bibinfo {author}
  {\bibfnamefont {J.}~\bibnamefont {Butet}}, \bibinfo {author} {\bibfnamefont
  {M.}~\bibnamefont {Kuittinen}}, \bibinfo {author} {\bibfnamefont {O.~J.~F.}\
  \bibnamefont {Martin}}, \ and\ \bibinfo {author} {\bibfnamefont
  {M.}~\bibnamefont {Kauranen}},\ }\href@noop {} {\bibfield  {journal}
  {\bibinfo  {journal} {Nano Lett.}\ }\textbf {\bibinfo {volume} {18}},\
  \bibinfo {pages} {7709} (\bibinfo {year} {2018})}\BibitemShut {NoStop}%
\bibitem [{\citenamefont {Yoon}\ \emph {et~al.}(2017)\citenamefont {Yoon},
  \citenamefont {Kim}, \citenamefont {So}, \citenamefont {Mun}, \citenamefont
  {Kim},\ and\ \citenamefont {Rho}}]{Yoon2017}%
  \BibitemOpen
  \bibfield  {author} {\bibinfo {author} {\bibfnamefont {G.}~\bibnamefont
  {Yoon}}, \bibinfo {author} {\bibfnamefont {I.}~\bibnamefont {Kim}}, \bibinfo
  {author} {\bibfnamefont {S.}~\bibnamefont {So}}, \bibinfo {author}
  {\bibfnamefont {J.}~\bibnamefont {Mun}}, \bibinfo {author} {\bibfnamefont
  {M.}~\bibnamefont {Kim}}, \ and\ \bibinfo {author} {\bibfnamefont
  {J.}~\bibnamefont {Rho}},\ }\href {\doibase 10.1038/s41598-017-06833-5}
  {\bibfield  {journal} {\bibinfo  {journal} {Sci. Rep.}\ }\textbf {\bibinfo
  {volume} {7}},\ \bibinfo {pages} {6668} (\bibinfo {year} {2017})}\BibitemShut
  {NoStop}%
\bibitem [{\citenamefont {Liu}\ \emph {et~al.}(2018)\citenamefont {Liu},
  \citenamefont {Wu}, \citenamefont {Zhang}, \citenamefont {Li}, \citenamefont
  {Zhang},\ and\ \citenamefont {Luo}}]{Liu2018}%
  \BibitemOpen
  \bibfield  {author} {\bibinfo {author} {\bibfnamefont {L.}~\bibnamefont
  {Liu}}, \bibinfo {author} {\bibfnamefont {L.}~\bibnamefont {Wu}}, \bibinfo
  {author} {\bibfnamefont {J.}~\bibnamefont {Zhang}}, \bibinfo {author}
  {\bibfnamefont {Z.}~\bibnamefont {Li}}, \bibinfo {author} {\bibfnamefont
  {B.}~\bibnamefont {Zhang}}, \ and\ \bibinfo {author} {\bibfnamefont
  {Y.}~\bibnamefont {Luo}},\ }\href {\doibase 10.1002/advs.201800661}
  {\bibfield  {journal} {\bibinfo  {journal} {Adv. Sci.}\ }\textbf {\bibinfo
  {volume} {5}},\ \bibinfo {pages} {1} (\bibinfo {year} {2018})}\BibitemShut
  {NoStop}%
\bibitem [{\citenamefont {Huttunen}\ \emph {et~al.}(2016)\citenamefont
  {Huttunen}, \citenamefont {Dolgaleva}, \citenamefont {T{\"{o}}rm{\"{a}}},\
  and\ \citenamefont {Boyd}}]{Huttunen2016a}%
  \BibitemOpen
  \bibfield  {author} {\bibinfo {author} {\bibfnamefont {M.~J.}\ \bibnamefont
  {Huttunen}}, \bibinfo {author} {\bibfnamefont {K.}~\bibnamefont {Dolgaleva}},
  \bibinfo {author} {\bibfnamefont {P.}~\bibnamefont {T{\"{o}}rm{\"{a}}}}, \
  and\ \bibinfo {author} {\bibfnamefont {R.~W.}\ \bibnamefont {Boyd}},\ }\href
  {\doibase 10.1364/OE.24.028279} {\bibfield  {journal} {\bibinfo  {journal}
  {Opt. Express}\ }\textbf {\bibinfo {volume} {24}},\ \bibinfo {pages} {28279}
  (\bibinfo {year} {2016})}\BibitemShut {NoStop}%
\bibitem [{\citenamefont {Segal}\ \emph {et~al.}(2015)\citenamefont {Segal},
  \citenamefont {Keren-Zur}, \citenamefont {Hendler},\ and\ \citenamefont
  {Ellenbogen}}]{Segal2015}%
  \BibitemOpen
  \bibfield  {author} {\bibinfo {author} {\bibfnamefont {N.}~\bibnamefont
  {Segal}}, \bibinfo {author} {\bibfnamefont {S.}~\bibnamefont {Keren-Zur}},
  \bibinfo {author} {\bibfnamefont {N.}~\bibnamefont {Hendler}}, \ and\
  \bibinfo {author} {\bibfnamefont {T.}~\bibnamefont {Ellenbogen}},\ }\href
  {\doibase 10.1038/nphoton.2015.17} {\bibfield  {journal} {\bibinfo  {journal}
  {Nat. Photonics}\ }\textbf {\bibinfo {volume} {9}},\ \bibinfo {pages} {180}
  (\bibinfo {year} {2015})}\BibitemShut {NoStop}%
\bibitem [{\citenamefont {Michaeli}\ \emph {et~al.}(2017)\citenamefont
  {Michaeli}, \citenamefont {Keren-Zur}, \citenamefont {Avayu}, \citenamefont
  {Suchowski},\ and\ \citenamefont {Ellenbogen}}]{Michaeli2017b}%
  \BibitemOpen
  \bibfield  {author} {\bibinfo {author} {\bibfnamefont {L.}~\bibnamefont
  {Michaeli}}, \bibinfo {author} {\bibfnamefont {S.}~\bibnamefont {Keren-Zur}},
  \bibinfo {author} {\bibfnamefont {O.}~\bibnamefont {Avayu}}, \bibinfo
  {author} {\bibfnamefont {H.}~\bibnamefont {Suchowski}}, \ and\ \bibinfo
  {author} {\bibfnamefont {T.}~\bibnamefont {Ellenbogen}},\ }\href {\doibase
  10.1103/PhysRevLett.118.243904} {\bibfield  {journal} {\bibinfo  {journal}
  {Phys. Rev. Lett.}\ }\textbf {\bibinfo {volume} {118}},\ \bibinfo {pages}
  {243904} (\bibinfo {year} {2017})}\BibitemShut {NoStop}%
\bibitem [{\citenamefont {Huttunen}\ \emph {et~al.}(2018)\citenamefont
  {Huttunen}, \citenamefont {Rasekh}, \citenamefont {Boyd},\ and\ \citenamefont
  {Dolgaleva}}]{Huttunen2018b}%
  \BibitemOpen
  \bibfield  {author} {\bibinfo {author} {\bibfnamefont {M.~J.}\ \bibnamefont
  {Huttunen}}, \bibinfo {author} {\bibfnamefont {P.}~\bibnamefont {Rasekh}},
  \bibinfo {author} {\bibfnamefont {R.~W.}\ \bibnamefont {Boyd}}, \ and\
  \bibinfo {author} {\bibfnamefont {K.}~\bibnamefont {Dolgaleva}},\ }\href
  {\doibase 10.1103/PhysRevA.97.053817} {\bibfield  {journal} {\bibinfo
  {journal} {Phys. Rev. A}\ }\textbf {\bibinfo {volume} {97}},\ \bibinfo
  {pages} {053817} (\bibinfo {year} {2018})}\BibitemShut {NoStop}%
\bibitem [{\citenamefont {Hooper}\ \emph {et~al.}(2018)\citenamefont {Hooper},
  \citenamefont {Kuppe}, \citenamefont {Wang}, \citenamefont {Wang},
  \citenamefont {Guan}, \citenamefont {Odom},\ and\ \citenamefont
  {Valev}}]{Hooper2018}%
  \BibitemOpen
  \bibfield  {author} {\bibinfo {author} {\bibfnamefont {D.~C.}\ \bibnamefont
  {Hooper}}, \bibinfo {author} {\bibfnamefont {C.}~\bibnamefont {Kuppe}},
  \bibinfo {author} {\bibfnamefont {D.}~\bibnamefont {Wang}}, \bibinfo {author}
  {\bibfnamefont {W.}~\bibnamefont {Wang}}, \bibinfo {author} {\bibfnamefont
  {J.}~\bibnamefont {Guan}}, \bibinfo {author} {\bibfnamefont {T.~W.}\
  \bibnamefont {Odom}}, \ and\ \bibinfo {author} {\bibfnamefont {V.~K.}\
  \bibnamefont {Valev}},\ }\href {\doibase 10.1021/acs.nanolett.8b03574}
  {\bibfield  {journal} {\bibinfo  {journal} {Nano Lett.}\ }\textbf {\bibinfo
  {volume} {19}},\ \bibinfo {pages} {165} (\bibinfo {year} {2018})}\BibitemShut
  {NoStop}%
\bibitem [{\citenamefont {Suchowski}\ \emph {et~al.}(2008)\citenamefont
  {Suchowski}, \citenamefont {Oron}, \citenamefont {Arie},\ and\ \citenamefont
  {Silberberg}}]{Suchowski2008}%
  \BibitemOpen
  \bibfield  {author} {\bibinfo {author} {\bibfnamefont {H.}~\bibnamefont
  {Suchowski}}, \bibinfo {author} {\bibfnamefont {D.}~\bibnamefont {Oron}},
  \bibinfo {author} {\bibfnamefont {A.}~\bibnamefont {Arie}}, \ and\ \bibinfo
  {author} {\bibfnamefont {Y.}~\bibnamefont {Silberberg}},\ }\href {\doibase
  10.1103/PhysRevA.78.063821} {\bibfield  {journal} {\bibinfo  {journal} {Phys.
  Rev. A - At. Mol. Opt. Phys.}\ }\textbf {\bibinfo {volume} {78}},\ \bibinfo
  {pages} {1} (\bibinfo {year} {2008})}\BibitemShut {NoStop}%
\bibitem [{\citenamefont {Suchowski}\ \emph {et~al.}(2009)\citenamefont
  {Suchowski}, \citenamefont {Prabhudesai}, \citenamefont {Oron}, \citenamefont
  {Arie},\ and\ \citenamefont {Silberberg}}]{Suchowski2009}%
  \BibitemOpen
  \bibfield  {author} {\bibinfo {author} {\bibfnamefont {H.}~\bibnamefont
  {Suchowski}}, \bibinfo {author} {\bibfnamefont {V.}~\bibnamefont
  {Prabhudesai}}, \bibinfo {author} {\bibfnamefont {D.}~\bibnamefont {Oron}},
  \bibinfo {author} {\bibfnamefont {A.}~\bibnamefont {Arie}}, \ and\ \bibinfo
  {author} {\bibfnamefont {Y.}~\bibnamefont {Silberberg}},\ }\href {\doibase
  10.1364/oe.17.012731} {\bibfield  {journal} {\bibinfo  {journal} {Opt.
  Express}\ }\textbf {\bibinfo {volume} {17}},\ \bibinfo {pages} {12731}
  (\bibinfo {year} {2009})}\BibitemShut {NoStop}%
\bibitem [{\citenamefont {Polyushkin}\ \emph {et~al.}(2011)\citenamefont
  {Polyushkin}, \citenamefont {Hendry}, \citenamefont {Stone},\ and\
  \citenamefont {Barnes}}]{Polyushkin2011}%
  \BibitemOpen
  \bibfield  {author} {\bibinfo {author} {\bibfnamefont {D.~K.}\ \bibnamefont
  {Polyushkin}}, \bibinfo {author} {\bibfnamefont {E.}~\bibnamefont {Hendry}},
  \bibinfo {author} {\bibfnamefont {E.~K.}\ \bibnamefont {Stone}}, \ and\
  \bibinfo {author} {\bibfnamefont {W.~L.}\ \bibnamefont {Barnes}},\
  }\href@noop {} {\bibfield  {journal} {\bibinfo  {journal} {Nano Lett.}\
  }\textbf {\bibinfo {volume} {11}},\ \bibinfo {pages} {4718} (\bibinfo {year}
  {2011})}\BibitemShut {NoStop}%
\bibitem [{\citenamefont {Luo}\ \emph {et~al.}(2014)\citenamefont {Luo},
  \citenamefont {Chatzakis}, \citenamefont {Wang}, \citenamefont {Niesler},
  \citenamefont {Wegener}, \citenamefont {Koschny},\ and\ \citenamefont
  {Soukoulis}}]{Luo2014}%
  \BibitemOpen
  \bibfield  {author} {\bibinfo {author} {\bibfnamefont {L.}~\bibnamefont
  {Luo}}, \bibinfo {author} {\bibfnamefont {I.}~\bibnamefont {Chatzakis}},
  \bibinfo {author} {\bibfnamefont {J.}~\bibnamefont {Wang}}, \bibinfo {author}
  {\bibfnamefont {F.~B.}\ \bibnamefont {Niesler}}, \bibinfo {author}
  {\bibfnamefont {M.}~\bibnamefont {Wegener}}, \bibinfo {author} {\bibfnamefont
  {T.}~\bibnamefont {Koschny}}, \ and\ \bibinfo {author} {\bibfnamefont
  {C.~M.}\ \bibnamefont {Soukoulis}},\ }\href {\doibase 10.1038/ncomms4055}
  {\bibfield  {journal} {\bibinfo  {journal} {Nat. Commun.}\ }\textbf {\bibinfo
  {volume} {5}},\ \bibinfo {pages} {1} (\bibinfo {year} {2014})}\BibitemShut
  {NoStop}%
\bibitem [{\citenamefont {Keren-Zur}\ \emph {et~al.}(2019)\citenamefont
  {Keren-Zur}, \citenamefont {Tal}, \citenamefont {Fleischer}, \citenamefont
  {Mittleman},\ and\ \citenamefont {Ellenbogen}}]{Keren-Zur2019a}%
  \BibitemOpen
  \bibfield  {author} {\bibinfo {author} {\bibfnamefont {S.}~\bibnamefont
  {Keren-Zur}}, \bibinfo {author} {\bibfnamefont {M.}~\bibnamefont {Tal}},
  \bibinfo {author} {\bibfnamefont {S.}~\bibnamefont {Fleischer}}, \bibinfo
  {author} {\bibfnamefont {D.~M.}\ \bibnamefont {Mittleman}}, \ and\ \bibinfo
  {author} {\bibfnamefont {T.}~\bibnamefont {Ellenbogen}},\ }\href {\doibase
  10.1038/s41467-019-09811-9} {\bibfield  {journal} {\bibinfo  {journal} {Nat.
  Commun.}\ }\textbf {\bibinfo {volume} {10}},\ \bibinfo {pages} {1778}
  (\bibinfo {year} {2019})}\BibitemShut {NoStop}%
\end{thebibliography}%

 %%%%%%% BEGINNING of SM %%%%%
 %%%%%%%%%% Merge with supplemental materials %%%%%%%%%%
\pagebreak

\widetext
\vspace{2cm}
\begin{center}
\textbf{\large Supplemental Material for \\
"Backward phase-matched second-harmonic generation from stacked metasurfaces"}
\end{center}
%%%%%%%%%% Merge with supplemental materials %%%%%%%%%%
%%%%%%%%%% Prefix a "S" to all equations, figures, tables and reset the counter %%%%%%%%%%
\setcounter{equation}{0}
\setcounter{figure}{0}
\setcounter{table}{0}
\setcounter{page}{1}
\makeatletter
\renewcommand{\theequation}{S\arabic{equation}}
\renewcommand{\thefigure}{S\arabic{figure}}
\renewcommand{\bibnumfmt}[1]{[#1]}
\renewcommand{\citenumfont}[1]{#1}
%%%%%%%%%% Prefix a "S" to all equations, figures, tables and reset the counter %%%%%%%%%%

\section{Transmission measurements}
\begin{figure}[h]
    \centering
    \includegraphics{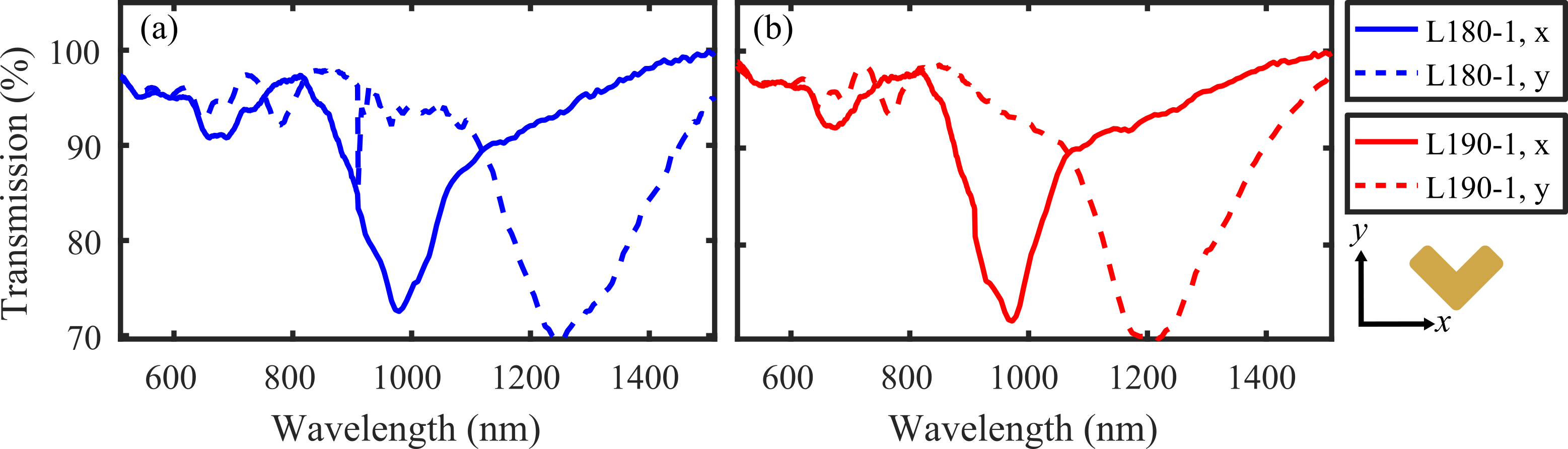}
    \caption{Transmission spectra of single layer devices (a) L180-1  (blue) and (b) L190-1 (red). Broad plasmon resonances are visible near 970 nm and 1200 nm for $y$-polarized (dashed) and $x$-polarized (solid) light, respectively.}
    \label{fig:lin_results}
\end{figure}

\noindent
In our Letter, we discuss the phase-engineering capabilities of our phase-matched metamaterial devices. These capabilities are governed by the localized surface plasmon resonances (LSPRs) of single metasurfaces. For the single metasurface devices L180-1 and L190-1, their transmission spectra (see Fig. \ref{fig:lin_results}) reveal the spectral locations of LSPRs. With $y$-polarized ($x$-polarized) incident light, the LSPRs peak near 975 nm (1230 nm) and 980 nm (1200 nm) for L180-1 and L190-1 devices, respectively. Close to these resonance wavelengths, light scattered from the metal nanoantenna exhibits phase changes, that allows to fulfill the phase-matching condition for devices made by stacking several metasurfaces on top of each other.

\section{Setup for measuring backward-emitted second-harmonic light}
\begin{figure}
    \centering
    \includegraphics{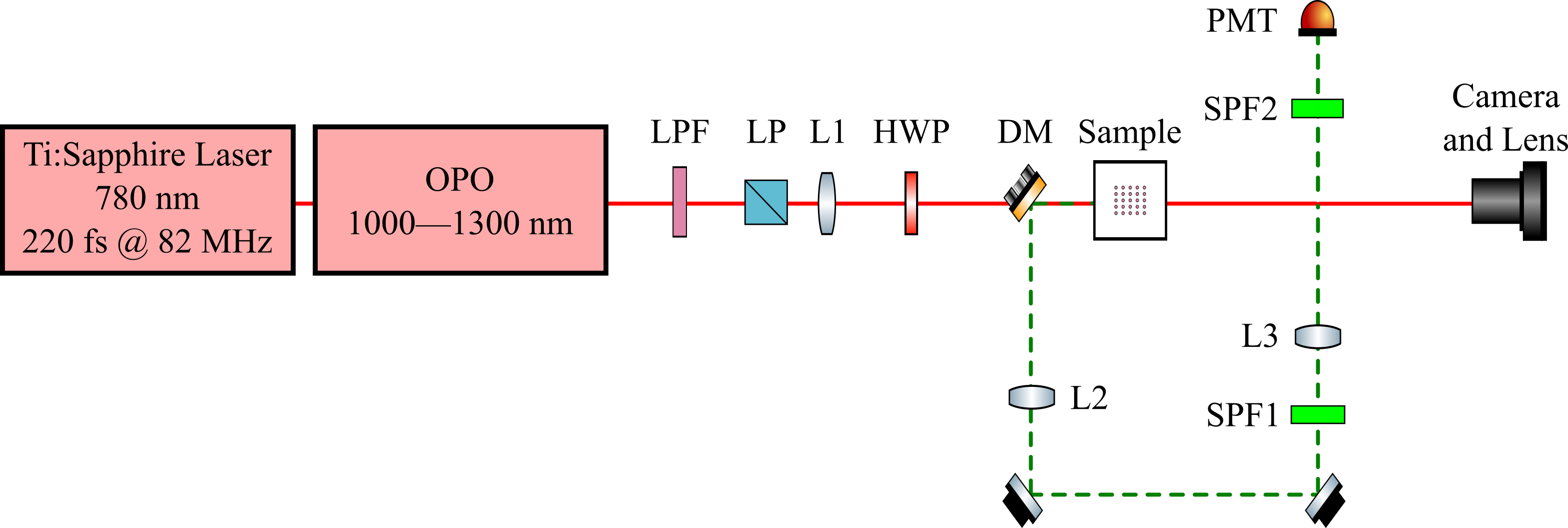}
    \caption{The setup used to measure SHG response of the sample. The setup consists of an optical parametric oscillator (OPO) pumped with a titanium sapphire femtosecond laser, long-pass filter (LPF), a linear polarizer (LP), a half-wave plate (HWP), lenses (L1, L2, and L3), dichroic mirror (DM), mirrors, an adjustable sample holder, short-pass filters (SPF1 and SPF2), a camera, and a photomultiplier tube (PMT).}
    \label{fig:setup}
\end{figure}

The second-harmonic (SH) emission from our devices was measured using the setup illustrated in Fig. \ref{fig:setup}, which is a modified version of the setup used in~\cite{Czaplicki2018}. We used an optical parametric oscillator pumped with a titanium sapphire femtosecond laser as a wavelength-tunable laser source. A long-pass filter ensured that only the correct wavelength range (1000--1300 nm) was guided to the sample.  Then, we used a linear polarizer to limit the power of the laser beam to 10 mW and set the polarization of the beam with a half-wave plate.  An achromatic lens was used to focus the laser on the sample that we imaged with a CMOS camera and a camera lens (MVL50M23). The back-propagating SHG emission was collected with another achromatic lens and guided to the reflection path (dashed green line in Fig. \ref{fig:setup}) with a dichroic mirror. The short-pass filter then ensured that the correct signal wavelengths were measured by the photomultiplier tube.

%%%%%%% END of SM %%%%%

\end{document}